\documentclass[prb,twocolumn]{revtex4}

\usepackage{amsmath}
\usepackage{amssymb}
\usepackage{psfrag}
\usepackage{pstricks}
\usepackage{graphicx}
\usepackage{subfigure}
\usepackage{dcolumn}
\usepackage{bm}

\def\summ{\sum\limits}
\def\be{\begin{equation}}
\def\ee{\end{equation}}
\def\nn{\nonumber}
\def\psid{\psi^{\dagger}}

\begin{document}

\bibliographystyle{prsty}

\title{Sagnac interference in Carbon nanotubes}

\author{Waheb Bishara}
\author{Gil Refael}
\author{Marc Bockrath}
\affiliation{ Department of Physics, California Institute of
  Technology, Pasadena, CA 91125}

\begin{abstract}
The Sagnac interference mode arises when two interfering
counterpropogating beams traverse a loop, but with their
velocities detuned by a small amount $2u$, with $v_{R/L}=v_F\pm u$. In this paper we perform a perturbative non-equilibrium calculation of
Sagnac interference in single channel wires as well as armchair nanotube loops. We study the dependence of the
Sagnac conductance oscillations on temperature and interactions. We
find that the Sagnac interference is not destroyed by strong
interactions, but becomes weakly dependent on the velocity detuning
$u$. In armchairs nanotubes with typical interaction strength,
$0.25 \leq g \leq 0.5$, we find that the necessary temperature for observing the
interference effect, $T_{SAG}$ is also only weakly dependent on the
interaction, and is enhanced by a factor of 8 relative to the temperature
necessary for observing Fabry-Perot interference in the same system, $T_{FP}$.

\end{abstract}
\maketitle

\section{Introduction}

\begin{figure}
  \includegraphics[width=2in]{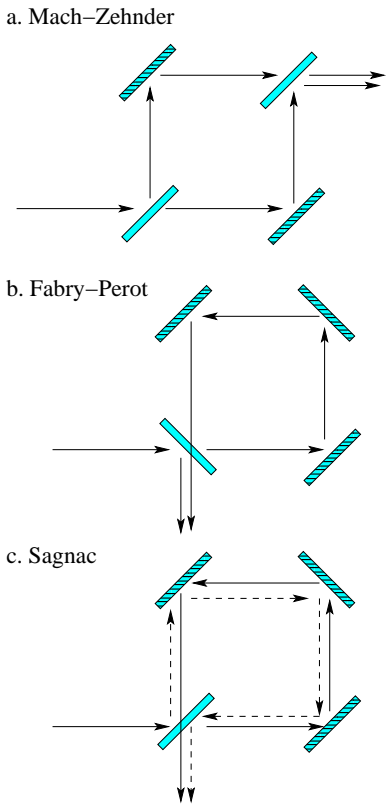}\\
  \caption{(a) In a Mach-Zehnder interferometer the input beam is
  split into two beams which traverse independent paths before being
  recombined. (b) In a Fabry-Perot interferometer a beam is split
  into a deflected ray, which is recombined at the output with a ray
  that traverses a loop. (c) The Sagnac interferometer splits the beam
  into a two beams which traverse the loop in two opposite
  orientations, and get recombined at the output. This allows a very
  sensitive measurement of the angular velocity of the interferometer,
  as it results in a different relative speed in the clockwise and
  couterclockwise rays. Clear rectangles
  represent beam-splitters, and patterned rectangles represent mirrors.  }\label{interferometer}
\end{figure}

One of the most tantalizing effects predicted by quantum mechanics is
the appearance of interference fringes when two {\it matter} beams
come together. These fringes provided the ultimate testimony to the
pertinence of quantum mechanics and the Schr\"odinger
equation. Interferometry of light is employed in many precision measurement devices. The
Mach-Zehnder interferometer produces interference between two beams
traversing two distinct paths, one of which passes through a test
chamber containing, for instance, a dilute gas (see Fig.\ref{interferometer}); this setup was
originally used to measure the refraction index of the gas in the
chamber. Fabry-Perot interferometer recombines a series of beams, where
the n'th beam traverses the optical path between two mirrors or through
a loop $n$ times. The narrowness of the resulting interference peaks
allows a precise measurement of a light beam's wave length, and is
commonly used to measure the Zeeman splitting of an atom in a magnetic
field. The most sensitive of all interference constructs, however, is
the Sagnac interference\cite{sagnac}. In this setup, a light beam is split into
two beams, which traverse the interferometer's loop both clock wise
and counter clock wise, before being recombined. In this case, the
interference fringes arise due to an absolute rotation, and provide
the most accurate measure of the angular velocity of the device. This
was used by Michelson to measure the absolute rotation of the
Earth. More recently, the Sagnac interference effect was cleverly used
to measure time reversal symmetry breaking in superconductors
\cite{XiaKapitulnik}.

Quantum mechanics opened the way for matter-wave interferometry.
Electron interferometry is a powerful probe of interaction effects on
low-energy phases of quantum matter, as demonstrated by numerous
examples. Mach-Zehnder interferometers reveal Aharonov-Bohm
oscillations and quantum hall effect edge
channels\cite{forster:075301,camino:155313,PhysRevB.67.033307,PhysRevLett.62.2523,buks},
and can probe exotic fractional quantum Hall
states\cite{feldman:186803,stern:016802,bonderson:016803}. Similarly,
two-path Mach-Zehnder interferometers can probe correlated states of quantum
dots\cite{sigrist:036805,PhysRevLett.74.4047}. Of particular interest
to us are metallic carbon nanotubes. The Luttinger liquid behavior in
these systems\cite{Bockrath99,KaneBalentsFisher97,YaoNature99} was partially verified through the observation of
Fabry-Perot interference in finite sections of the nano-tube
\cite{LiangNature01}. The Fabry-Perot interference should, in principle,
allow the observation of spin-charge
separation and determination of the interaction parameters of the
Luttinger liquid \cite{PecaBalents}. But the similar
energy scales of the spin and charge modes' interference patterns has
made such experimental observation challenging.

\begin{figure}
  \includegraphics[width=2in]{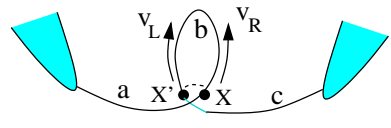}\\
  \caption{Schematic of a nanotube loop reported in Ref. \onlinecite{RefaelSagnac}. This geometry allows electrons to tunnel from the point X on the loop to a distant point X' on the other end of the loop, and vice versa. We refer to this process as cross-loop tunneling. An electron entering from the left can traverse the loop moving right with velocity $v_R$, without scattering, or tunnel from X to X' and traverse the loop moving left with velocity $v_L$.}\label{knot}
\end{figure}

\begin{figure}
  \includegraphics[width=3in]{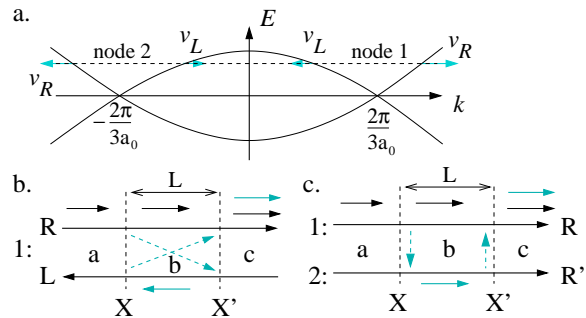}\\
  \caption{(a) The energy spectrum of an armchair nanotube. When the chemical potential is tuned away from the degeneracy points by a gate voltage, the left and right movers in each node will have different velocities, which leads to the Sagnac interference in the loop geometry. (b) The scatterings we consider in this paper tunnel, say, a right moving electron from a given node at point $X$ to a left moving electron, of the same node, at point $X'$, and vice versa. (c) Sagnac interference can also arise without the loop geometry through inter-node tunneling, since right movers at node $2$ have the same velocity as left movers at node $1$.}
  \label{Spectrum}
\end{figure}

The most sensitive interferometer of all, however, the Sagnac
interferometer, has not been seriously explored yet in the context of
interacting electronic systems. In Ref. \onlinecite{RefaelSagnac} we
proposed that this interference naturally occurs in metallic armchair
nanotube loops (Fig. \ref{knot}). Instead of rotation, the Sagnac interference arises due to the band
velocity difference between right- and left-moving electrons about
each Dirac node. This
velocity difference is present whenever the electronic Fermi surface is
tuned away from the Dirac points at half-filling, as shown
in Fig. \ref{Spectrum}(a).  The operating principle of the electronic Sagnac
effect has the same origin as the universal conductance fluctuations,
and weak-localization effects in disordered two-dimensional electron
gases \cite{PhysRevLett.42.673,Gorkov,AltshulerAronov}. In nanotubes, it can also appear due to band-scattering in a
pair of impurities \cite{Jiang2003}.

Because the Sagnac effect involves electrons traversing
the same path in two different directions, rather than repeating the
same path as in Fabry-Perot interference, the phase accumulation is
extremely small. Therefore Sagnac interference exhibits large-period conductance fluctuations as a function of gate-
and source-drain voltages, and is expected to persist to high temperatures in
comparison to Fabry-Perot interference, which is more
sensitive to thermal dephasing. This interference mode should thus be
able to reveal much more precise information about the unique
state of interacting electrons in thin quantum wires.

Our goal in this manuscript is to thoroughly explore the range and
robustness of the Sagnac interference mode, concentrating on armchair
Carbon nanotubes. The questions we will ask concern the amplitude
of this interference mode as a function of the temperature, gate and
source-drain voltage, and Luttinger parameter of the nanotube.

The paper is organized as follows. In Section \ref{SingleChannel}, as
a warm-up, we analyze the simpler case of Sagnac interference in a
single channel of right- and left-moving electrons. In \ref{Model1ch} we introduce the model
of a single channel with a linearized spectrum, and the cross-loop
tunneling which will give rise to the Sagnac interference. In
\ref{Keldysh} and \ref{PerturbativeCalculation} we set up the
non-equilibrium perturbative calculation of the conductance in the
presence of cross-loop tunneling, and in \ref{Volt} and \ref{Temp} we
analyze the behavior of the oscillating conductance as a function of
gate and bias voltages and temperature. In Section \ref{Nanotubes} we
repeat the above steps for the physically relevant case of Carbon
nanotubes, including spin and node degeneracies in the calculation,
and remark on the similarities and differences from the single channel
case. Finally we conclude with a discussion of the experimental
implications of our calculations.

\section{Sagnac interference in a single channel} \label{SingleChannel}

As discussed in the introduction, the Sagnac interference in the loop
geometry is due the the asymmetry between the velocities of the left
and right moving electrons. To demonstrate this in the simplest form,
we first study in this section a single channel with a single type of
left and right movers. In a carbon nanotube, there will be four such
channels due to spin and node degeneracies.

\subsection{The Model}\label{Model1ch}

We start with a single one dimensional channel of electrons and a
linearized spectrum, with different left and right mover
velocities, and a density-density interaction. The Hamiltonian density
for this system is:

\begin{align}\label{Hamiltonian1ch}
{\cal H}_{1ch}& = -i\hbar v_{R} \psid_{R}\partial_x\psi_{R}+i \hbar v_{L} \psid_{L}\partial_x\psi_{R} \nn \\& +\lambda \left(\psid_{R}\psi_{R}+\psid_{L}\psi_{L}\right)^2
\end{align}
where the operator $\psi^{\dagger}_{R/L}$ creates a right/left moving electron, with the velocity:
\be
v_{R/L}=v_F \pm u.
\ee
The scattering we are interested in is the one which takes a right
moving electron at one side of the loop, point $X$ is Figure
\ref{knot}, and scatters it to a left moving electron at the other
side of the loop, point $X'$, and vice versa. This process has been
dubbed Cross-Loop scattering in Ref. [\onlinecite{RefaelSagnac}]. The
same effect can also be obtained without the loop geometry by
inter-node tunneling \cite{Jiang2003}, since right movers at node $1$ move with the same velocity as left movers at node $2$. This inter-node tunneling is shown in Figure \ref{Spectrum}(c). If we choose our coordinate along the loop such that the point $X$ corresponds to $x=0$ and the point $X'$ corresponds to $x=L$, then this scattering process is described by the Hamiltonian:
\begin{align}\label{Hbs1ch}
H_{bs}=\Gamma_1 \psid_{R}(0)\psi_{L}(L) + h.c.   \nn\\
 +\Gamma_2\psid_{L}(0)\psi_{R}(L) +h.c.
\end{align}

In the presence of the quartic density-density interactions in the Hamiltonian, Eq. (\ref{Hamiltonian1ch}), it is useful to use the standard bosonization procedure , since the Hamiltonian is quadratic in terms of the bosonic fields. The electron fields are bosonized as follows:
\be
\psi_{R/L}\sim e^{i(\phi\pm\theta)}
\ee

where $\theta$ and $\phi$ are bosonic fields that satisfy the
commutation relations $[\theta(x),\phi(x')]=i(\pi/2)sgn(x-x')$; also,
the total density and the current density are given by
$\frac{1}{\pi}\nabla\theta=\rho_R+\rho_L$, and
$\frac{1}{\pi}\nabla\phi=\rho_R-\rho_L$, respectively. The Hamiltonian in terms of the bosonic fields becomes:
\be \label{H1ch}
H_{1ch}=\frac{\hbar v}{2\pi}\int dx \left[ \frac{1}{g}(\nabla \theta)^2 + g (\nabla \phi)^2 +2\frac{u}{v}\nabla \theta \nabla \phi \right]
\ee
where $g=\left( 1+\frac{2\lambda}{\pi\hbar v_F}\right) ^{-1/2}$ is the Luttinger interaction parameter and $v=v_F/g$. This is the familiar Hamiltonian of a 1D interacting electron system, with the addition of the $u$ term which gives left and right moving particles different velocities. Indeed, this Hamiltonian can be easily diagonalized and the left and right velocities turn out to be for a general value of the interaction parameter $g$:
\be \label{vg_pm_u}
v_{R/L}=v\pm u = \frac{v_F}{g} \pm u.
\ee

Our goal is to calculate the effects of the Sagnac interference
as seen in the conductance as a function of the applied bias and
gate voltages, and as a function of temperature. Due to the applied
voltages the system is not in equilibrium, and we must turn to the
Keldysh non-equilibrium formalism \cite{KeldyshRef,RammerRMP}. Below we carry
out this analysis first for the simplified electron gas with the scattering
Hamiltonian $H_{bs}$, Eq. (\ref{Hbs1ch}), as a perturbation.

\subsection{Non-Equilibrium correlation functions and conductance}\label{Keldysh}

The response of the loop to a bias source-drain voltage can be
analyzed using the non-equilibrium Keldysh
formalism. Following Ref.\onlinecite{PecaBalents}, we assume that in
the distant past, before turning on the backscattering, the left and
right moving electrons separately had well defined thermal
distributions set by separate chemical potentials. The density matrix
corresponding to this initial distribution at temperature $T=1/\beta$
is:
\be
\hat{\rho}_{V}=\frac{1}{Z_{V}}e^{-\beta \hat{H}_{V}},
\ee
with $Z_{V}=Tr[e^{-\beta \hat{H}_{V}}]$ and the Hamiltonian which takes into account the applied voltages is:
\begin{align}
H_{V}=H_{1ch} - e \frac{V_{sd}}{2}\left(N_R-N_L\right)  -  \alpha e V_g\left(N_R+N_L\right) \nn \\
=H_{1ch} - e \frac{V_{sd}}{2} \int dx  \frac{\nabla \phi}{\pi} -\alpha e V_g  \int dx  \frac{\nabla \theta}{\pi}
\end{align}

The gate voltage, $V_g$, simply couples to the total charge density,
with $\alpha$ being a geometrical factor of the system, while the
source-drain voltage, $V_{sd}$, induces the imbalance in the chemical
potentials of the left and right movers.

As explained in Ref. \onlinecite{PecaBalents}, both $V_{sd}$ and
$V_{g}$ can be eliminated from the unperturbed action by an
appropriate unitary transformation, which is equivalent to shifting
the bosonic fields by a function of space and time; this is easy
to see if one writes down the Lagrangian including the
voltages\cite{RefaelSagnac}. The equivalent shifts for the case at
hand are:
\begin{align}
\theta&\rightarrow \theta+\frac{\alpha g^2 e V_g}{\hbar
  v_F}\frac{1}{1-g^2u^2/v_F^2} x-\frac{eV_{sd}}{2\hbar} t, \nn \\
\phi &\rightarrow \phi- \frac{\alpha g^2 e V_g}{\hbar
  v_F}\frac{u/v_F}{1-g^2u^2/v_F^2} x,
\label{shifts}
 \end{align}
These shifts remove the voltages from the Hamiltonian $H_V$ and
therefore all the correlations to appear in the calculation will be
equilibrium correlation functions with respect to $H_{1ch}$. The
dependance on the applied voltages now appears in the scattering
Hamiltonian, $H_{bs}$, due to the shifted bosonic fields.

Let us now focus our attention at the charge current, which in
the bosonic language is $\hat{I}=(e/\pi)\partial_t \theta$. After
performing the unitary transformation described above we can write the
formal expression  for the expectation value of the current in the
usual interaction picture \cite{PecaBalents}:
\be \label{Expectation}
\langle I \rangle = I_0 + \frac{1}{Z_{V=0}} Tr \left( e^{-\beta H_{1ch}} \hat{T}_K \left\{ \hat{I}_K(x,t) e^{-i\int_{\cal C} dt' H'_{bs}(t')}\right\} \right)
\ee

\begin{figure}
  \includegraphics[width=3in]{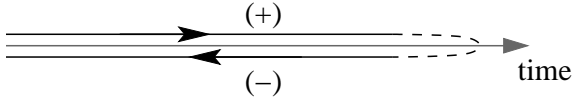}\\
  \caption{The Keldysh contour used in the non-equilibrium calculation. The Keldysh time ordering operator $T_K$ orders operators along the contour, so fields on the (+) branch is always at an earlier time than fields on the (-) branch.}\label{CK}
\end{figure}
$\hat{T}_K$ is the time ordering operator along the Keldysh contour
shown in Fig.\ref{CK}, and $\hat{I}_K(x,t)$ is the symmetrized current
operator with respect to the two branches of the contour. The current
$I_0=e^2V_{sd}/h$ is the ideal current that would flow in the absence
of backscattering in a completely transmitting channel, and it
explicitly appears due to the shift of the $\theta$ field. The
Hamiltonian $H'_{bs}$ denotes the scattering Hamiltonian
$H_{bs}$ with the properly shifted bosonic fields. The expression for
the current can be expanded in powers of $H'_{bs}$, and all the
correlation functions to appear in this expansion are equilibrium
correlation functions at temperature $1/\beta$. If we denote by
$\theta^+$ and $\theta^-$ the fields on the forward branch and
backward branch of the Keldysh contour respectively, then time
ordering along the contour means that $\theta^+\theta^+$ correlations
have the usual time ordering, $\theta^-\theta^-$ are anti-time
ordered, and $\theta^+(t)$ is always earlier in time that
$\theta^-(t')$. The same applies for all the fields.

It is useful to apply a Keldysh rotation to the fields, $\theta^\pm=\theta\pm \frac{i}{2}\tilde{\theta}$, and similarly for $\phi$. The correlation function $\langle T_K \tilde{\theta}(t)\tilde{\theta}(t') \rangle$ vanishes by construction, and we define :
\begin{align}
C^{\theta}(x,t;x',t')=&\langle T_K  \theta(x,t)\theta(x',t')\rangle \nn \\
= &\frac{1}{2}\langle \{ \hat{\theta}(x,t),\hat{\theta}(x',t') \} \rangle \nn\\
R^{\theta}(x,t;x',t')=&\langle T_K \theta(x,t)\tilde{\theta}(x',t') \rangle \nn \\
=& -i\Theta(t-t')\langle [ \hat{\theta}(x,t),\hat{\theta}(x',t')]\rangle
\end{align}
and similarly for the $\phi_j$ fields, and for the mixed correlations:
\begin{align} \label{Ctp}
C^{\theta\phi}(x,t;x',t')=&\langle T_K  \theta(x,t)\phi(x',t')\rangle \nn \\
= &\frac{1}{2}\langle \{ \hat{\theta}(x,t),\hat{\phi}(x',t') \} \rangle \nn\\
R^{\theta\phi}(x,t;x',t')=&\langle T_K \theta(x,t)\tilde{\phi}(x',t') \rangle \nn \\
=& -i\Theta(t-t')\langle [ \hat{\theta}(x,t),\hat{\phi}(x',t')]\rangle
\end{align}
where operators with a hat are simply the time dependent operators with no time ordering. As explained above, these correlation function are to be evaluated in equilibrium, and therefore are easily explicitly calculated (Appendix
\ref{corrs}). Due to translational invariance in time and space,
these correlations are functions of $x-x'$ and $t-t'$, for example:
\begin{align}
C^{\theta\phi}(x,t) \equiv  C^{\theta\phi}(x,t;0,0) = \nn  \\
\frac{1}{4}\left[  \log \left(v_{R}\sinh\left(\frac{(v_{R}t-x)\pi}{\beta v_{R}}\right)\right) \right. \nn \\
-  \left.   \log \left(v_{L}\sinh\left(\frac{(v_{L}t+x)\pi}{\beta v_{L}}\right)\right) \right].
\end{align}

\subsection{Perturbation Theory} \label{PerturbativeCalculation}

The Sagnac interference fringes occur already with weak bakcscattering
at the base of the loop, and can be deduced from a perturbation
analysis of the tunneling Hamiltonian, Eq. (\ref{Hbs1ch}). As outlined
above, to calculate the current,
$I_{1ch}=\langle \frac{e}{\pi}\partial_t \theta \rangle$, we absorb the gate and bias
voltages, $V_g$ and $V_{sd}$ respectively, in the shifts in Eq.\ref{shifts}, which allow us to move the
voltages from the unperturbed Hamiltonian $H_{1ch}$ to the
backscattering perturbation, $H_{bs}$. Then, we expand the formal
expression we found for the current using the Keldysh technique,
Eq. \ref{Expectation}, in powers of $H_{bs}$, and use Wick's theorem
to evaluate the resulting contributions.


To lowest nontrivial order, which is second order in $H_{bs}$, we
obtain after a lengthy calculation:
\be
I_{1ch}=\frac{e^2 V_{sd}}{h} + I_{co} + I_{inco}
\label{Itotal}
\ee
The first term is simply the current that would flow through the system in the absence of backscattering. The coherent current, $I_{co}$, oscillates with the gate voltages $V_g$, and is given by:
\begin{align}\label{Ico}
I_{co}= &  c \, \Gamma_1\Gamma_2 \, \cos\left(\frac{2u g^2 L\alpha}{\hbar^2 v_F^2(1-g^2u^2/v_F^2)}V_g\right) \times \nn \\
 & \int dt \,\sin(\frac{eV_{sd}}{\hbar}t) \,e^{-C_{co}(L,t)} \sin (R_{co}(L,t))
\end{align}
where c is a constant of order unity, and we assume that $\Gamma_i$ are real for simplicity. The incoherent current, $I_{inco}$, is independent of the gate voltage, and is given by:
\begin{align}
\label{Iinco}
&I_{inco}=  \nn \\
& c \,\left( \,\Gamma_1^2 \sum_{\eta=\pm} \int dt \,\sin(\frac{eV_{sd}}{\hbar}t) \,e^{-C_{inco}^{\eta}(L,t)} \sin (R_{inco}(L,t))  \right)  \nn \\
& +c \, \left( \Gamma_1 \rightarrow \Gamma_2, L \rightarrow -L \right)
\end{align}

The functions $C_{co}$, $C_{inco}^{\pm}$, $R_{co}$ and $R_{inco}$ are
complicated combinations of the correlation functions defined in
section \ref{Keldysh} and are given explicitly in Appendix
\ref{corrs}. These functions do not simplify, partly due to the fact
that the correlation functions in this problem are not symmetric under
$x\rightarrow -x$ since left and right movers have different velocities.

\subsection{Voltage dependence of the single-mode
  Sagnac interference} \label{Volt}

The voltage current characteristics given in Eqs. (\ref{Itotal} -
\ref{Iinco}) can be evaluated numerically to obtain the voltage and
temperature dependence of the single-mode Sagnac interference.
The period
of the interference as a function of the gate voltage ($I_{co}$) are easily
observed to be (for small $u/v_F$):
\be
\Delta V_g^{Sagnac} \approx \frac{v_F}{u} \frac{\pi \hbar^2 v_F} {\alpha g^2 L} = \frac{v_F}{u} \Delta V_g^{FP}
\ee
where $\Delta V_g^{FP}$ is the period in gate voltage for Fabry-Perot
interference. Fabry-Perot interference occurs whenever part of the
wave's trajectory can be repeated. Since the Sagnac interference
involves traversing the same path in two different directions, the
phase difference accumulated in the process is much smaller than the
difference incurred by repeating part of the path, and therefore the
period of the Sagnac interference is much
larger than the period of the Fabry-Perot interference. Such large period oscillations have been
experimentally observed in Carbon nanotubes, in the loop geometry, as
reported in Ref.\onlinecite{RefaelSagnac}, in addition to the shorter
period Fabry-Perot oscillations.

\begin{figure}
  \includegraphics[width=3.5in]{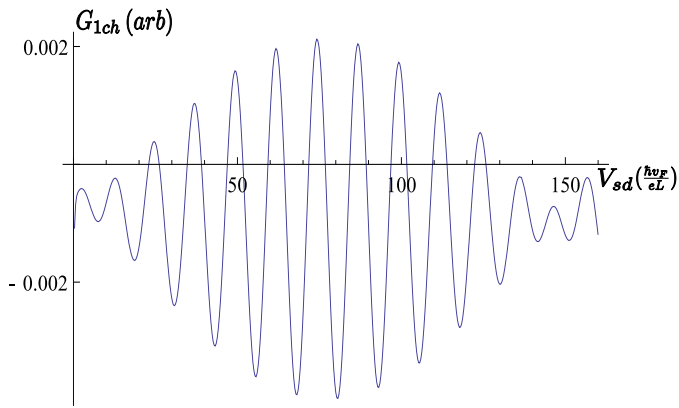}
  \caption{Differential conductance oscillations of a single channel of fermions, as a function of bias
    voltage $V_{sd}$, for velocity detuning $u/v_F=0.1$ and interaction strength
    $g=0.5$. The beating is due to the only two voltage oscillation frequencies in the
    problem, $\Omega_1=\frac{eL}{\hbar(v+u)}$ and $\Omega_2=\frac{eL}{\hbar(v-u)}$,
    where $v=v_F/g$. The voltage is in units of $\hbar v_F/eL$. The shorter voltage oscillation periods is $\Delta V_{sd}= 2\pi \left( \frac{\Omega_1+\Omega_2}{2}\right)^{-1}\approx 12.5 \, \hbar v_F/eL$, and the large oscillation period is  $\Delta V_{sd}= 2\pi \left( \frac{\Omega_1-\Omega_2}{2}\right)^{-1}\approx 250 \,\hbar v_F/eL$. } \label{beating}
\end{figure}

For a given gate voltage, both the coherent and incoherent parts of
the current oscillate with the bias voltage $V_{sd}$. This oscillation is due to the fact that in the presence of bias voltage, the Fermi energy of the left- and right- moving electrons are different by $V_{sd}$, and hence their Fermi wavevectors are different also and they would acquire different phases traversing the loop. This oscillation will be present even for no velocity detuning, $u=0$. When the detuning is finite, $u\neq 0$, the differential conductance $G_{1ch}=\partial I_{1ch}/\partial V_{sd}$
will show a beating pattern due to the two different left and right
moving excitation velocities. Here we are only considering the Sagnac oscillations arising from the
cross-loop tunneling, Eq. (\ref{Hbs1ch}). Figure \ref{beating} shows
the oscillations of the differential conductance at a fixed gate
voltage. For non-interacting electrons, the beating pattern
corresponds to the addition of two harmonics with two different
frequencies in voltage, $\sin ( \Omega_R V_{sd})$ and $\sin ( \Omega_L V_{sd})$, with $\Omega_{R/L}=\frac{eL}{\hbar v_{R/L}}$ and $v_{R/L}=v_F\pm u$. The beating pattern will then display fast oscillations with voltage period $\Delta V_{sd}^{fast}=2\pi \left(\frac{|\Omega_R+\Omega_L|}{2}\right)^{-1}$, and slow voltage oscillations with period $\Delta V_{sd}^{slow}=2\pi \left(\frac{\Omega_R-\Omega_L}{2}\right)^{-1}$. For interacting fermions, $g\neq 1$, the oscillations will
not be simple harmonic oscillations. Still, the periods will be
evident and will have the same functional form, in terms of $v_{R/L}$, as the frequencies in the non-interacting case. The periods $\Delta V_{sd}^{fast}$ and $\Delta V_{sd}^{slow}$ do depend on $g$ through velocities $v_L$ and $v_R$, $v_{R/L}=\frac{v_F}{g}\pm u$ (Eq. \ref{vg_pm_u}).

These oscillation, generally, lie atop a powerlaw behavior of the differential conductance as a function
of $V_{sd}$, as expected from the known behavior of the conductance in
the presence of impurity backscattering \cite{KaneFisherPRB92}. For
backscattering from an impurity in a Luttinger liquid, the
backscattered current, for low temperature, behaves as $I\propto
V_{sd}^{2g-1}$. We chose to plot the Sagnac oscillations as a function
of $V_{sd}$ (Fig. \ref{beating}) for the interaction parameter $g=0.5$
since for that value the corresponding power law would be $I\propto
V_{sd}^0$, and the contribution of such a powerlaw to the {\it
  differential} conductance would vanish, making the oscillation atop
this powerlaw more visible.

\subsection{Temperature dependence of the single-mode
  Sagnac interference} \label{Temp}

Next we consider the temperature dependence of the
gate-voltage driven oscillations in the coherent part of the current. As argued in
Ref. \onlinecite{RefaelSagnac}, the large period Sagnac oscillations are
expected to be observed at much higher temperature than the shorter
period Fabry-Perot oscillations. This difference in temperature
behavior can be easily understood by examining the phase giving rise
to the interference in both cases. In the Fabry-Perot case for a loop, the lowest
order interference is between a beam of electrons which is not
scattered, and a beam of electrons which, due to scattering at the
base of the loop, does
a roundtrip between the the two scattering points. The phase
difference between these two beams at energy $E$ is $\Delta
\phi_{FP}=k_RL=\frac{1}{v_R}L\frac{E}{\hbar}$.
Finite temperature effectively causes uncertainty of order $T$ in the
energy $E$, and the interference pattern will be washed out when the
uncertainty of the the phase $\Delta \phi_{FP}$ is of order $2\pi$,
which happens at a temperature $T_{FP}=\frac{2\pi\hbar}{L}v_R$.

In the Sagnac case, the
interference is between a beam that traverses the loop moving left and
one which traverses the loop moving right. The phase difference
between these two beams at energy $E$ is $\Delta \phi_{SAG}=k_L L -
k_R L = \left(\frac{1}{v_L}-\frac{1}{v_R}\right)L\frac{E}{\hbar}$, and
this interference will be washed out at temperature
$T_{SAG}=\frac{2\pi\hbar}{L} \left(
\frac{1}{v_L}-\frac{1}{v_R}\right)^{-1}$. For non-interacting
electrons the right and left moving velocities are $v_{R/L}=v_F\pm
u$. Thus to lowest order in $u/v_F$, the highest temperatures for
observing interference according to the argument above are:
 \be
 T_{FP}\approx \frac{\pi\hbar v_F}{L}; \;\;\; T_{SAG}\approx\frac{\pi\hbar v_F}{L} \frac{v_F}{u}=T_{FP}\cdot \frac{v_F}{u}
 \ee
 For non-interacting electrons, we expect the Sagnac interference to survive to a temperature higher by a factor of $v_F/u$ than the corresponding Fabry-Perot temperature. We will show through explicit calculation that this is indeed true for the non-interacting case. For interacting electrons, we will see that $T_{SAG}$ will still be considerably larger than $T_{FP}$, but their ratio is less than the dramatic $v_F/u$ ratio.

\begin{figure}
  \includegraphics[width=3in]{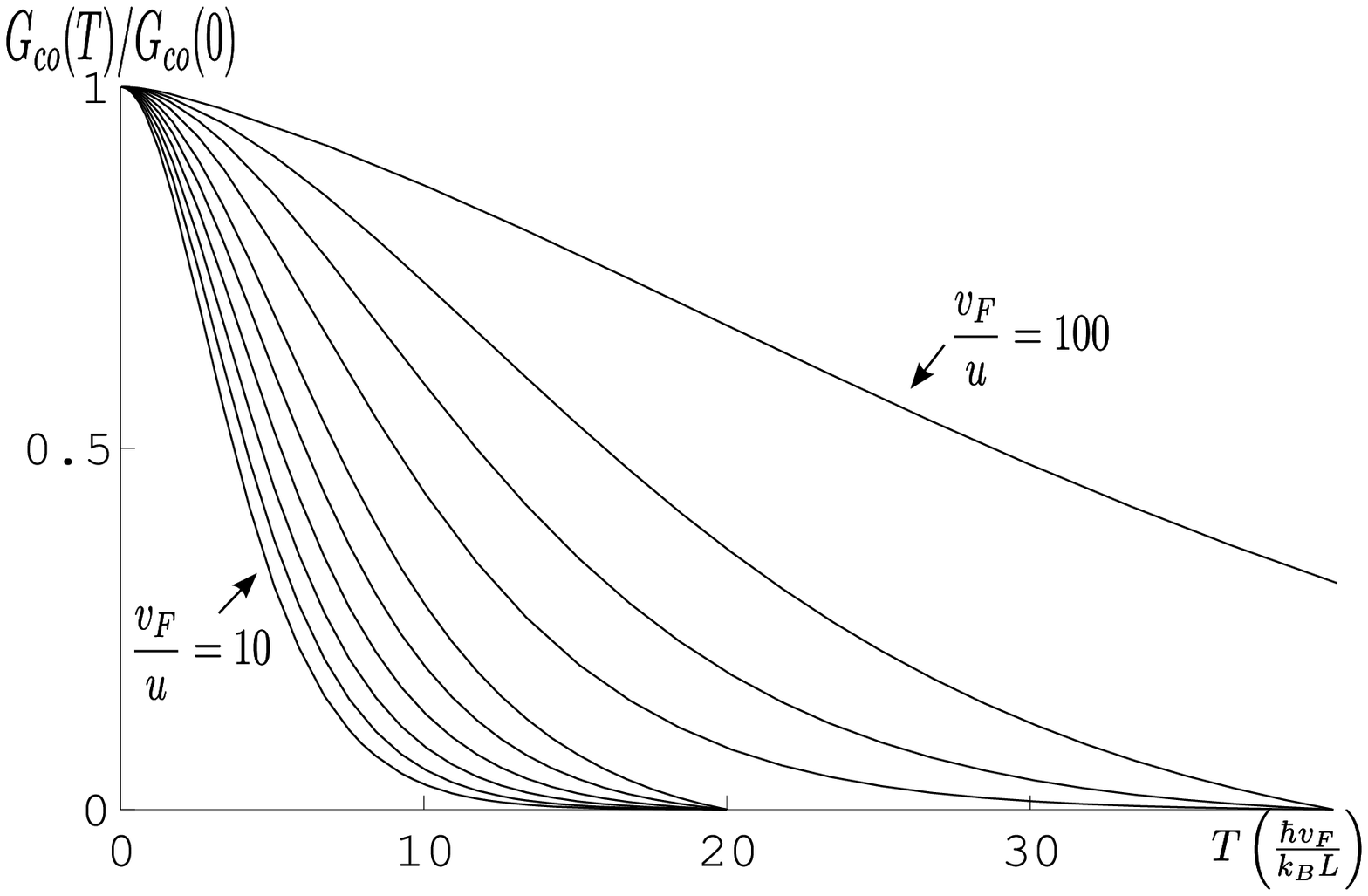}\\
  \caption{Coherent Sagnac oscillation amplitude vs. temperature for
  different values of $(v_F/u)$, for noninteracting electrons ($g=1$). The slowest decaying plot corresponds to $v_F/u=100$, and the fastest decaying plot corresponds to $v_F/u$=10. Temperature is given in units of $\hbar v_F/k_B L$.}\label{g1}
\end{figure}

 To explore the Sagnac temperature range, we evaluate the amplitude of the coherent oscillations (the oscillations in $V_g$) as a function of
temperature, for different interaction parameters $g$ and different
ratios of $u/v_F$. For non-interacting electrons, $g=1$, we find that the Sagnac
oscillations indeed survive up to a high temperature, which is a factor of
$v_F/u$ higher than the corresponding Fabry-Perot oscillations. Figure
\ref{g1} plots the oscillation amplitude as a function of
temperature, normalized by its zero-temperature value, and for different
values of $u/v_F$. The functional dependence on temperature is given approximately by:
\be
\frac{G_{co}(T)}{G_{co}(T=0)}=\left(\frac{2\pi k_B L T}{\hbar v_F}\right) \left( \frac{u}{v_F}\right)\frac{1}{\sinh (2\pi k_B T \frac{L}{\hbar v_F}\frac{u}{v_F})}
\ee
This result is similar to the exact form of the temperature dependence of the
Fabry-Perot interference amplitude \cite{RecherEtAl}, with the only
difference being the factor of $u/v_F$. Therefore, the Sagnac
oscillations of non-interacting electrons indeed survive up to a
temperatures which are a factor of  $v_F/u$ larger than the Fabry-Perot oscillations.

\begin{figure}
  \includegraphics[width=3.5in]{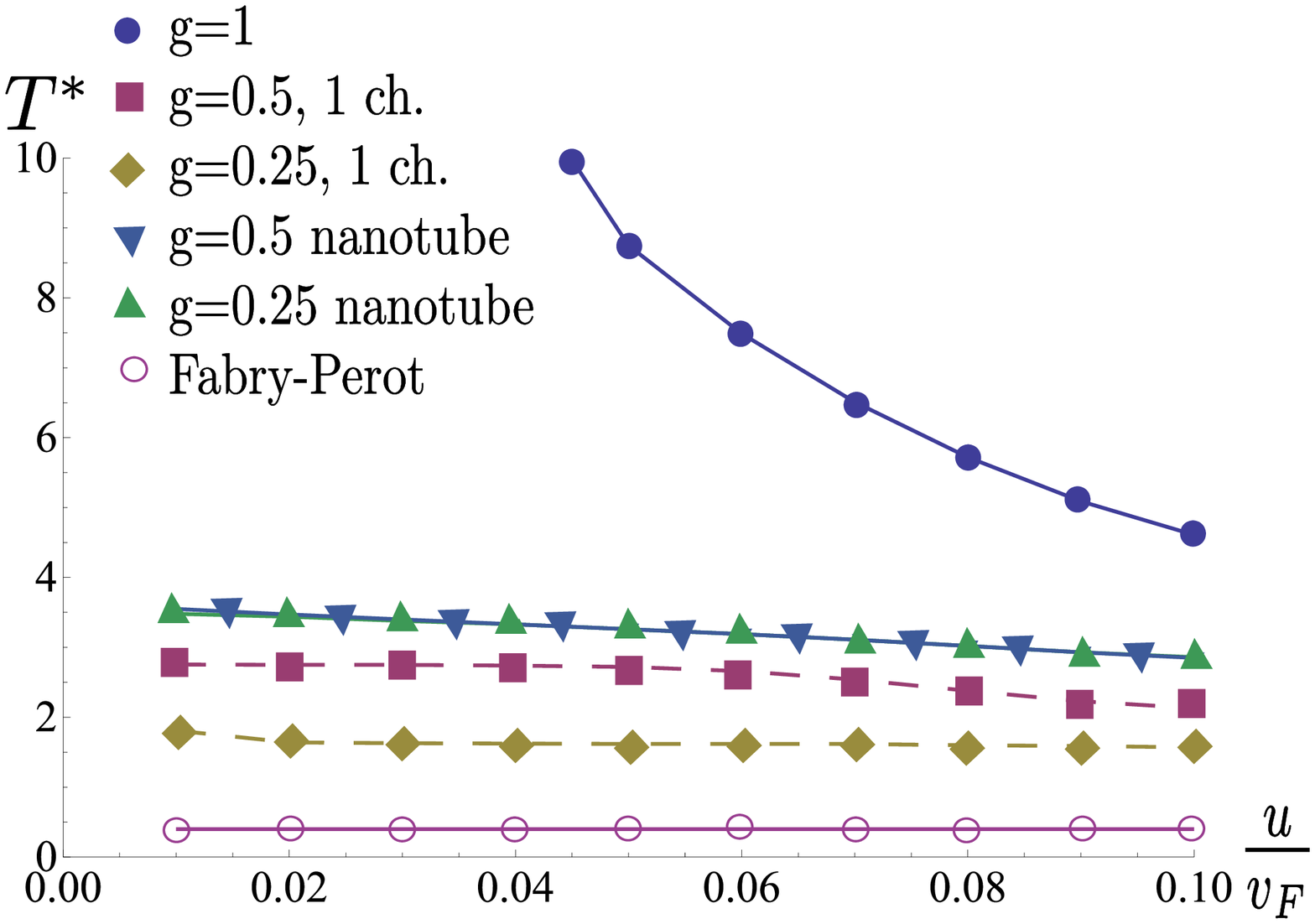}\\
  \caption{$T^*$ vs. $u/v_F$, where $T^*$ is the temperature at which
  the coherent differential conductance (the part of the conductance
  which oscillates with gate voltage) reaches $e^{-1}$ of its zero
  temperature value. For a non-interacting system, $g=1$, the single
  channel case gives the same temperature dependence as the case with
  spin and node degeneracies, $T^*\propto v_F/u$. The single
  channel temperature dependence is given for $g=0.5$ (squares,
  dashed), and $g=0.25$ (diamonds, dashed). The Carbon nanotube
  temperature dependence is given for $g=0.5$ (triangles), and
  $g=0.25$ (inverted triangles). Temperature is given in units of
  $\hbar v_F/k_B L$. For reference, the $T^*$ corresponding to the $g=1$ Fabry-Perot oscillations is also plotted.}\label{TvsU}
\end{figure}

For interacting electrons, $g\neq 1$, the Sagnac interference still
survives up to temperatures significantly higher than the corresponding
Fabry-Perot temperature scales, but the
enhancement is suppressed compared to that of non-interacting
electrons. Figure \ref{TvsU} shows the Sagnac temperature scale $T^*$
vs. $u/v_F$ for three different values of the interaction parameter
$g$, where we define $T^*$ to be the temperature at which the
amplitude of the oscillations reaches $e^{-1}$ of its amplitude at
zero temperature. For
non-interacting electrons $T^*$ is strongly dependent on the ratio
$u/v_F$ as discussed above. For the interaction parameter values
$g=0.5$ and $g=0.25$ (Dashed lines), the temperature $T^*$ is only
weakly dependent on the ratio $u/v_F$. As an example for the resulting enhancement of the Sagnac
compared to the Fabry Perot interference, consider  $g=0.25$, where
the $T^*$ temperature scale for the Sagnac oscillations is roughly $1.6 \, \hbar v_F/k_B L$, a factor of 4 enhancement over $T^*$ of the non-interacting Fabry-Perot oscillations which is $0.42 \, \hbar v_F/k_B L$, despite the suppression of the Sagnac $T^*$ due to interactions. As can be seen in the figure, for $g=0.5$ the enhancement is about 7. While it is difficult to extract the analytic dependence of the temperature on the interaction parameter, one can repeat our
calculation for any value of $g$.

\section{Interference in Nanotubes} \label{Nanotubes}

Equipped with our understanding of the single-channel Sagnac
interference, we can now consider the likely physical system where it
may be observed: a metallic Carbon nanotube with four different Dirac nodes. We now
add the spin and node degeneracies of a Carbon nanotube, and examine
their effect on the Sagnac interference pattern voltage and
temperature dependence.

\subsection{The Model}
The energy spectrum of a Carbon nanotube is shown in figure \ref{Spectrum}
a. This spectrum is usually linearized around the Fermi surface, which
yields four chiral modes, two left moving and two right moving (not
including spin), with linear dispersion. These modes can be bosonized
and treated within the Luttinger Liquid theory framework, as we have
done in the single channel case in the previous sections. All these
modes are usually assumed to have the same velocity, the Fermi
velocity $v_F$. For the purposes of this paper, it is important to
notice that when the Fermi surface is away from the degeneracy points
where the upper and lower bands meet, linearizing the spectrum
actually gives two different velocities which we shall note
$v_{\pm}=v_F \pm u$. The linearized Hamiltonian density is, then:
\begin{align}\label{Hamiltonian}
{\cal H}_{4ch}& =i \sum_{a=1}^2 \sum_{\sigma=\uparrow,\downarrow} \left(v_{Ra} \psid_{Ra\sigma}\partial_x\psi_{Ra\sigma}-v_{La} \psid_{La\sigma}\partial_x\psi_{Ra\sigma}\right) \nn \\& +\lambda \left[ \sum_{a=1}^2\sum_{\sigma=\uparrow,\downarrow}\left(\psid_{Ra\sigma}\psi_{Ra\sigma}+\psid_{La\sigma}\psi_{La\sigma}\right)\right]^2
\end{align}
where $\psi_{R/La\alpha}$ stands for a right/left moving electron at node $a$ with spin $\sigma$, and we added a total charge density interaction term. The velocities that appear in the Hamiltonian are:
\begin{align}\label{vpm}
v_{R/L1\sigma}=v_F \pm u = v_{\pm}\nn \\
v_{R/L2\sigma}=v_F \mp u = v_{\mp}.
\end{align}
Thus for $u>0$, at node $1$ right movers are faster than left
movers, while at node $2$ the opposite is true. Now, the nonlinearity
of the elctronic spectrum in a Carbon nanotube needs to be taken into
account when considering the velocity
difference, $u$; it depends on the detuning of the chemical potential away
from the degeneracy points.

The scattering process we are interested
in is very similar to the one we had in the single channel case. We
need to consider a term that scatters a right mover at one end of the loop to a left mover at the
other end of the loop, conserving spin and node quantum numbers,
\begin{align}\label{Hbs}
H_{bs}=\sum\limits_{\sigma,a=1,2}\left[\Gamma_1 \psid_{Ra\sigma}(0)\psi_{La\sigma}(L) + h.c. \right.  \nn\\
\left. +\Gamma_2\psid_{La\sigma}(0)\psi_{Ra\sigma}(L) +h.c. \right].
\end{align}

Next we bosonize the electron field operators in the nanotube. The slowly oscillating parts can be written as:
\be
\psi_{R/La\sigma}\sim e^{i(\phi_{a\sigma}\pm\theta_{a\sigma})}
\ee
where $\theta_{a\sigma}$ and $\phi_{a\sigma}$ are bosonic fields that satisfy the commutation relations $[\theta_{a\sigma}(x),\phi_{a'\sigma'}(x')]=i(\pi/2)\delta_{a,a'}\delta_{\sigma,\sigma'}sgn(x-x')$. The Hamiltonian in terms of the bosonic fields is \cite{RefaelSagnac}:
\be
\begin{array}{ll}
& H_{4ch}= \frac{\hbar v_F}{2\pi}\summ_{\sigma,a=1,2}\int dx \left[\left(\nabla\theta_{a\sigma}\right)^2+\left(\nabla\phi_{a\sigma}\right)^2\right.\\ & \left. +(-1)^{a+1} 2\frac{u}{v_F}\nabla\phi_{a\sigma}\nabla\theta_{a\sigma}\right]
+\int dx \lambda\left(\summ_{\sigma,a=1,\,2}\frac{1}{\pi}\nabla\theta^{\sigma}_a\right)^2.
\end{array}
\label{HB}
\ee
If the velocities of all branches of the spectrum were equal, i.e. $u=0$, then the Hamiltonian $H_B$
would be diagonalized by the spin and node symmetric and antisymmetric
combinations of the $\theta$'s and $\phi$'s
\cite{KaneBalentsFisher97}. By diagonalizing we mean a linear mapping
of the $\phi$ and $\theta$ fields such that the
Hamiltonian takes the form of four independent channels, each resembling of $H_{1ch}$, Eq. (\ref{H1ch}). When $u\neq 0$, there still
exists a local transformation
$\theta_{a\sigma}=\sum_{j=1..4}\left(A^j_{a\sigma}\theta_j+B^j_{a\sigma}\phi_j\right)$
that diagonalizes the Hamiltonian, but it is a more complicated
combination of the fields that depends on $u$ and $\lambda$, and mixes
the $theta$ and $\phi$ fields, which makes the conductance calculation
quite cumbersome. While the details of this transformation are given in appendix
\ref{Diagonalization}, the diagonal Hamiltonian is:
\begin{align} \label{HB2}
 H_{4ch}=
  \summ_{j=3,4} \frac{\hbar v_j}{2\pi} \int dx & \left[\frac{1}{g_j}\left(\nabla\theta_j\right)^2+g_j\left(\nabla\phi_{j}\right)^2\right] \nn  \\ + \summ_{j=1,2}\frac{\hbar v_F}{2\pi} \int dx & \left[\left(\nabla\theta_j\right)^2+\left(\nabla\phi_{j}\right)^2 \right. &  \nn \\ &  \left.  +(-1)^{j+1} 2\frac{u}{v_F}\nabla\phi_{j}\nabla\theta_j\right].
\end{align}
The fields $\theta_{1/2}$ and $\phi_{1/2}$ are the spin antisymmetric
combinations of $\theta_{1/2\sigma}$ and $\phi_{1/2\sigma}$
respectively. Since the interaction term in Eq.(\ref{Hamiltonian})
involves only the spin symmetric combinations, the spin antisymmetric
combinations are untouched and still have the left and right moving
velocities as in Eq.(\ref{vpm}). On the other hand, the fields
$\theta_{3/4}$ and $\phi_{3/4}$ are not simply the remaining symmetric
combination and mix the remaining $\theta$'s and $\phi$'s. These
fields have the same left and right moving velocity, which is:
\be\label{velocities}
v_{3/4}=\frac{v_F}{\sqrt{2}}\sqrt{1+\frac{1}{g^2}+2\frac{u^2}{v_F^2}\pm\sqrt{\left(1-\frac{1}{g^2}\right)^2+8\frac{u^2}{v_F^2}\left(1+\frac{1}{g^2}\right)}}
\ee
where $g=\left( 1+\frac{8\lambda}{\pi\hbar v_F}\right) ^{-1/2}$ is the
Luttinger parameter.\cite{Oreg95}

Fortunately, for the region of parameters which is of interest, namely
strong interactions, $g\le0.5$ and $u/v_F\le 0.1$, the exact change of
basis required to diagonalize the spin symmetric part of the
Hamiltonian is very close to the usual node symmetric/antisymetric
basis. This can be explicitly seen, for example, from the velocities
of these modes. For this entire range of parameters, the velocities of
the diagonal fields, given by Eq.(\ref{velocities}), are at most $1\%$
different from the values we expect for the left-right symmetric
system, which are $v_F/g$ and $v_F$. Due to the strong interactions in
this spin symmetric sector, the velocity asymmetry is unimportant, and
it is for this reason that we choose to still use the node
symmetric-antisymmetric basis and treat these fields as the diagonal
ones. In Appendix \ref{Diagonalization} we elaborate on and justify
this approximation. Note that the velocity asymmetry is still apparent in the
non-interacting spin antisymmetric modes labeled by $j=1$ and $j=2$ in
Eq. (\ref{HB2}).

\subsection{Perturbation Theory}

Using the diagonal form of the Hamiltonian with the above
approximation, we proceed to calculate the current,
$I=(e/\pi)\langle \sum_{a\sigma}\partial_t \theta_{a\sigma}\rangle$,
as in Section \ref{SingleChannel}. The applied voltages now couple to
the total density and total number of left movers and right movers:
\begin{align}
H_{V}=H_{4ch} - e \frac{V_{sd}}{2}\left(N_R-N_L\right)  -  \alpha e V_g\left(N_R+N_L\right) \nn \\
=H_{4ch} - e \frac{V_{sd}}{2} \int dx \sum_{\sigma,a} \frac{\nabla \phi_{a\sigma}}{\pi} -e V_g  \int dx \sum_{\sigma,a} \frac{\nabla \theta_{a\sigma}}{\pi}
\end{align}
The external voltages can be removed from the Hamiltonian by the appropriate shift of the bosonic fields:
 \begin{align}
\theta_{a\sigma}&\rightarrow \theta_{a\sigma}+\frac{\alpha g^2 e V_g}{\hbar
  v_F}\frac{1}{1-g^2u^2/v_F^2} x-\frac{eV_{sd}}{2\hbar} t, \nn \\
\phi_{a\sigma} &\rightarrow \phi_{a\sigma}+(-1)^a \frac{\alpha g^2 e V_g}{\hbar
  v_F}\frac{u/v_F}{1-g^2u^2/v_F^2} x.
\label{shifts2}
 \end{align}

Again we use the Keldysh contour to write the formal expression for
the current, as in Eq. (\ref{Expectation}), and expand it to lowest
order in the appropriate $H'_{bs}$ which contains the voltage
dependence due to the shifts of the fields. The approximation we made
above, namely that it is the node symmetric/antisymmetric combination
which diagonalize the Hamiltonian, allows us to write the current in a
very similar form to the single channel case:
\be \label{I4ch}
I_{4ch}=4 \frac{e^2 V_{sd}}{h} + \tilde{I}_{co} + \tilde{I}_{inco}
\ee

The first term on the right hand side of Eq. (\ref{I4ch} is  the
current that would flow in the nanotube in the absence of
backscattering. The second term is the coherent current which
oscillates with the gate voltage:
\begin{align}\label{Ico4ch}
\tilde{I}_{co}= &  c \, \Gamma_1\Gamma_2 \, \cos\left(\frac{2u g^2 L\alpha}{\hbar^2 v_F^2(1-g^2u^2/v_F^2)}V_g\right) \times \nn \\
 & \int dt \,\sin(\frac{eV_{sd}}{\hbar}t) \,e^{-\tilde{C}_{co}(L,t)} \sin (\tilde{R}_{co}(L,t))
\end{align}
and the third term is the incoherent current, which is independent of the gate voltage:
\begin{align}
&\tilde{I}_{inco}=  \nn \\
& c \,\left( \,\Gamma_1^2 \sum_{\eta=\pm} \int dt \,\sin(\frac{eV_{sd}}{\hbar}t) \,e^{-\tilde{C}_{inco}^{\eta}
(L,t)} \sin (\tilde{R}_{inco}(L,t))  \right) + \nn \\
& c \, \left( \Gamma_1 \rightarrow \Gamma_2, L \rightarrow -L \right)
\end{align}

The function $\tilde{C}_{co}$, $\tilde{C}_{inco}^{\pm}$, $\tilde{R}_{co}$
and $\tilde{R}_{inco}$ are related to the single channel correlation
functions as explained in Appendix \ref{corrs}.
\begin{figure}
  \includegraphics[width=3.5in]{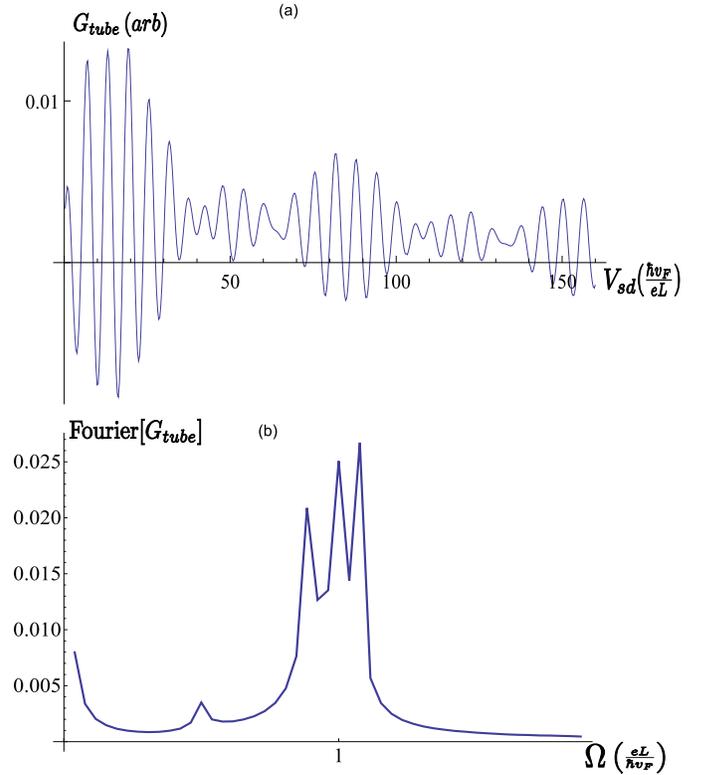}\\
  \caption{(a) Differential conductance oscillations for a nanotube, i.e. including both spins and both nodes in the spectrum, for velocity detuning $u/v_F=0.1$, and interaction strength $g=0.5$. In the nanotube case, the beating is due to the four voltage frequencies in the problem, $\Omega_i=\frac{eL}{\hbar v_i}$, where $v_{1/2}=v_F\pm u$, $v_3\approx v_F$ and $v_4\approx v_F/g$. (b) The voltage Fourier transform of the oscillations in (a) clearly displays the four dominant frequencies , $\Omega_{1-4}$, corresponding to the four velocities in the problem, and encode the nanotube parameters $v_F$, $g$ and $u/v_F$. }\label{beating_tube}
\end{figure}

\subsection{Temperature and Voltage Dependence in Carbon Nanotubes}

As in the single channel case, we find there is a coherent part of the
interference current which oscillates as a function of the gate
voltage with a large period, much larger than the Fabry-Perot
oscillation period, as seen explicitly from the voltage dependence of
$\tilde{I}_{co}$.

The differential conductance  $\partial I_{4ch}/\partial V_{sd}$, on
the other hand,  displays a beating pattern, but a more complicated one
than in the single-channel case, since there are four different velocities in the
problem now: $v_F\pm u$, $v_3\approx v_F$ and $v_4\approx
v_F/g$. Figure \ref{beating_tube} shows the differential conductance  of the nanotube, $\partial I_{4ch}/\partial
V_{sd}$, and its Fourier transform. From the Fourier analysis we see
that clearly there are four dominant frequencies, which correspond to the
four different velocities of the collective modes in the
nanotube. Thus a careful observation of the large-period, and robust,
Sangac interference allows, in principle, to extract the
nanotube parameters, namely the interaction strength $g$ and the
velocity mismatch $u$ from the Fourier transform of the conductance as a function of bias voltage, up to temperatures much higher than the Fabry-Perot oscillations temperatures.

The temperature dependence of the Sagnac interference in the nanotube
case is qualitatively similar to the single channel case. In the
absence of interactions ($g=1$), the interference can be observed to
the scale $T^*$ proportional to $v_F/u$; in the
presence of strong interactions, however, $T^*$ becomes only weakly dependent
on $u$. Unlike the single channel case, $T^*$ in the nanotube case is
also only very weakly dependent on $g$ in the range $g\leq 0.5$. This is
due to the fact that only one of the four modes which diagonalize the
Hamiltonian are interacting and depend on $g$. For the same reason,
$T^*$ is higher in the case of the nanotube than in the single channel
case, i.e. the reduction of $T^*$ due to interactions is not as severe
in the nanotube case. The temperature dependence on $u$ and $g$ is
plotted in Figure \ref{TvsU}. In the range mentioned above, $T^*_{SAG}\approx 2.8 \, \frac{\hbar v_F}{k_B L} \approx 7 \, T^{*}_{FP}$.

\section{Summary and Conclusions}\textbf{}

In this paper we investigated the conductance
oscillations in carbon nanotubes due to Sagnac interference. In
addition to theoretical interest in this large-period interference
mode, the motivation for our study also comes from a recent experimental realization of carbon nanotube loops
\cite{RefaelSagnac}. The same interference mode can arise also without the loop
geometry in the presence of internode backscattering in the nanotube,
as pointed out in Ref. \onlinecite{Jiang2003}.

The source of the Sagnac conductance oscillations is the difference in
the velocities of left and right moving excitations in a carbon
nanotube when the chemical potential is tuned away from half
filling. Compared to the more familiar Fabry-Perot oscillations
\cite{LiangNature01}, Sagnac oscillations are expected to have a much larger period in
gate voltage, and, as we show, in non-interacting wires survive to a temperature a factor of
$v_F/u$ higher than that required to observe Fabry-Perot
oscillations.

In interacting electronic wires,  the above
temperature estimation for free fermions does not apply. Our results for a
single channel Luttinger liquid are that $T_{SAG}$ becomes only
weakly dependent on $v_F/u$, although still strongly dependent on
$g$. From our $g=0.5,\,0.25$ results, the enhancement of relative to
the FP interference is roughly: $T_{SAG}\approx 15 g\, T_{FP}$ in the
range $u/v_F<0.1$.

For a strongly
interacting armchair nanotube, $g\le 0.5$, we find that $T_{SAG}$
becomes not only
weakly dependent on $v_F/u$, but also nearly independent of $g$. The
Sagnac interference is expected to survive upto $T^*_{SAG}\sim 3
\frac{\hbar v_F}{k_B L}\approx 7-8  T^{*}_{FP}$. Considering that Fabry-Perot oscillations have been observed in
nanotubes up to $T=10K$ \cite{LiangNature01}, Sagnac oscillations
should be observed up to about $70K$ in nanotubes, despite the strong
interactions.

There is also something to be learnt from examining the behavior of
the conductance as a function of the applied voltages. We saw
that Sagnac oscillations would have a large period of oscillations in
the applied gate voltage $V_g$; this period itself is a function of the
gate voltage, through the dependence of the velocity difference $v_R-v_L=2 u$. Using typical values of a
nanotube parameters (e.g. Ref. \onlinecite{RefaelSagnac}), $v_F=8\cdot 10^5
m/s$, $L=7\mu m$, $g=0.3$ and $\alpha=1/30$, the period of oscillation
in the gate voltage would be $\Delta V_g= \frac{2\pi \hbar
  v_F}{eL}\frac{1}{\alpha g^2}\frac{v_F}{u}\approx 17 V$, consistent
with the observed oscillations in Ref. \onlinecite{RefaelSagnac}.

On the other hand, oscillations of the conductance as a function of the
applied bias voltage $V_{sd}$ depend not only on the bare velocities,
but also on the interaction strength. A Fourier transform of the
Sagnac oscillations as a function of $V_{sd}$, we show, contains four
different frequencies corresponding to the four different velocities
in the problem, which are roughly  $v_F\pm u$,$v_F$ and $v_F/g$. Using the same
parameters as above we get $\Delta V_{sd}=\frac{2\pi\hbar
  v_F}{eL}\approx 0.5 mV$. This period is much smaller than the
bandwidth of a nanotube which is a few eV, so in principle many
oscillation periods can be observed and the longer period oscillations
should also be measurable, allowing the slower frequency oscillations
to appear in the Fourier transform. Observation of these frequencies
would allow us to read off the parameters of the nanotube, $v_F$, $u$
and $g$, at temperatures up to $T^*_{SAG}\approx 70mK$, which is higher than the temperatures associated with Fabry-Perot oscillations.

In the single channel case, for non-interacting electrons, we were able to extract an analytic expression for the temperature behavior of the conductance gate voltage oscillations :
\be
\frac{G_{co}(T)}{G_{co}(T=0)}=\left(\frac{2\pi k_B L T}{\hbar v_F}\right) \left( \frac{u}{v_F}\right)\frac{1}{\sinh (2\pi k_B T \frac{L}{\hbar v_F}\frac{u}{v_F})}
\ee
and it is apparent how the ratio $v_F/u$ directly enters the temperature scale. Unfortunately, we were so far unable to extract analytic expression
for $T_{SAG}$ in terms of $g$ and $u/v_F$ for the interacting single channel or interacting nanotube cases, inspite of the progress on
the qualitative understanding our numerical results allow. Such an analytical understanding should be
the focus of a future effort.

As can be observed in Figs. \ref{interferometer} and \ref{knot}, the paths
giving rise to the Sagnac intereference are similar to the paths that
give rise to weak localization phenomena in 2d disordered
conductors. In this work we also essentially show that even in the
presence of strong interactions, the interference survives. It is
tempting to extrapolate from our results that weak localization should
also survive strong interactions. This, however, is presumably true so
long that scattering events are dominated by small momentum
transfer. Nevertheless, our results suggest that a Luttinger liquid
with charge and spin modes will still exhibit weak-localization
effects, but suppressed, and only
weakly dependent on the detuning between counter propogating
electrons. Therefore the magnetoresistance should also be strongly
suppressed at low fields.

\acknowledgements

We are indebted to Jinseong Hu, Chetan Nayak, Yuval Oreg, Leonid Pryadko, and Jan
von Delft for illuminating discussions. MB is grateful for support by the ONR.

\appendix
\section{Diagonalizing the Hamiltonian with Node and Spin Degeneracies}\label{Diagonalization}

In this appendix we show how to diagonalize the Hamiltonian $H_{4ch}$ of
Eq. (\ref{HB}), where diagonalizing entails finding the appropriate
change of basis that will transform $H_{4ch}$ to the sum of four
Hamiltonians, each having a form resembling the single channel $H_{1ch}$
of Eq. (\ref{H1ch}). We also explain here the approximations we have used in our
calculation.

The first step in the diagonalization of $H_B$ is to change the basis
from the spin up/down to the spin symmetric/antisymmetic basis at each node:
\be
\theta_{a\pm}=\frac{\theta_{a\uparrow}\pm \theta_{a,\downarrow}}{\sqrt{2}}
\ee
applying the same transformation to the $\phi$'s as well.
We notice that the density-density interaction term involves only the
spin symmetric fields $\theta_{a+}$, hence the spin antisymmetric
fields decouple and appear as two non-interacting ($g=1$) copies of
the single channel problem, described by the Hamiltonian $H_{1ch}$,
with right moving velocity of $v_F\pm u$ and left moving velocities of
$v_F \mp u$. These are the fields labeled with $j=1$ and $j=2$ in
Eq. (\ref{HB2}).

The Hamiltonian for the spin symmetric fields has a similar form to
our starting point Hamiltonian, $H_{4ch}$:
\begin{align}
&H_{+}= \frac{\hbar v_F}{2\pi}\summ_{a=1,2}\int dx \left[\left(\nabla\theta_{a+}\right)^2+\left(\nabla\phi_{a+}\right)^2\right.\\
& \left. +(-1)^{a+1} 2\frac{u}{v_F}\nabla\phi_{a+}\nabla\theta_{a+}\right]
+\int dx \, 2\lambda\left(\summ_{a=1,\,2}\frac{1}{\pi}\nabla\theta_{a+}\right)^2 \nn
\end{align}

In the absence of the $u$ term, $H_+$ is easily diagonalized by taking the node symmetric and antisymmetric combinations of the fields:
\be \label{SpinSym}
\theta_{3/4}=\frac{\theta_{1+} \pm \theta_{2+}}{\sqrt{2}}
\ee
The resulting diagonal Hamiltonian would be:
\begin{align} \label{u0H}
\left. H_{+}\right|_{u=0}=& \frac{\hbar v_3}{2\pi}\int dx \left[ \frac{1}{g_3}(\nabla \theta_3)^2 + g_3 (\nabla \phi_3)^2 \right] \nn \\
+& \frac{\hbar v_4}{2\pi}\int dx  \left[\frac{1}{g_4} (\nabla \theta_4)^2 + g_4  (\nabla \phi_4)^2 \right]
\end{align}
with $v_3=v_F$, $v_4=v_F/g$, $g_3 =1$ and $g_4=g$.

When we consider
$u\neq 0$, it is still possible to apply a $g$ and $u$ dependent transformation to the fields, that
will restore $H_+$ to the form in Eq. (\ref{u0H}), with
velocities $v_{3/4}$ given by Eq. (\ref{velocities}). The field mixing
this transformation entails, however, considerably complicates the book keeping in our
perturbative calculation. Fortunately, we can show that a good
approximation is to simply set $u$ to zero in $H_+$
when the interactions are strong, and simply use the transformation
given by Eq. (\ref{SpinSym}). The first indication that this
approximation is valid is that the
exact velocities $v_{3/4}$ differ from the $u=0$ velocities
by at most $1\%$ in the entire range of parameters we are interested
in, which is $u/v_F \leq 0.1$ and $g \leq 0.5$

Another indication that this
approximation is valid comes from the analysis of the single channel
problem in Section
\ref{SingleChannel}. In the single channel case we derived exact
expressions for the Sagnac interference, and
found that for $g=0.5$ and $g=0.25$, the temperature dependence is
only weakly dependent on $u/v_F$; furthermore $u$ only enters directly
in the expression for the oscillation period of the conductance as a function of gate voltage,
the dependence we have explicitly in our expression for the coherent
current $I_{co}$, Eq. \ref{Ico}.

Finally, we can also calculate the exact combination
of fields that diagonalizes $H_+$, and verify that indeed they are
very close to the node symmetric/antisymmetric combinations for the
range of $g$ and $u$ of interest. As an example, the explicit change of basis from the node symmetric/antisymmetric basis to the diagonalizing basis for $g=1/2$, to second order in $u/v_F$, is:
\begin{align}
\mathbb{I}_{4x4} +
\left(
\begin{array}{cccc}
-\frac{71}{144} (\frac{u}{v_F})^2 & 0 & 0 & \frac{2\sqrt{2}}{3} ( \frac{u}{v_F}) \\
0 & -\frac{89}{144} (\frac{u}{v_F})^2 & \frac{5}{3\sqrt{2}} ( \frac{u}{v_F})& 0 \\
0 &  \frac{-2\sqrt{2}}{3} ( \frac{u}{v_F}) & -\frac{29}{36} (\frac{u}{v_F})^2 & 0\\
-\frac{5}{3\sqrt{2}} ( \frac{u}{v_F}) & 0 & 0 & -\frac{11}{36} (\frac{u}{v_F})^2
\end{array} \right)
\end{align}

We see that the is matrix is close to the identity matrix $\mathbb{I}_{4x4}$, since $\frac{u}{v_F} \ll 1$. The deviation from the identity becomes even smaller for smaller $g$. Note that for $g\approx 1$ the corresponding change of basis matrix is not close to the identity matrix and our approximation fails.

We stress that setting $u$ to zero
in $H_+$ is simply a good numerical approximation which simplifies the calculation, and not equivalent
to setting $u$ to zero in the entire problem, as $u$ still appears in
spin anti-symmetric part of the Hamiltonian (where $g=1$), and also in
the gate voltage dependence.

\section{Correlation functions}\label{corrs}

Let us now connect the explicit expressions for the coherent and
incoherent currents given in Section \ref{PerturbativeCalculation} and
Section \ref{Nanotubes}, Equations (\ref{Ico}) and (\ref{Iinco}), using the correlation functions defined in
Section \ref{Keldysh}.

It is useful to define the following combination of $C^{\theta}$:
\be
\overline{C^{\theta}(x,t)}=C^\theta(0,0)-C^\theta(x,t)
\ee
and similarly for $\overline{C^{\phi}}$.

In the single channel case discussed in Section \ref{SingleChannel}, there are only a single $\theta$ field and a single $\phi$ field, with the Hamiltonian given by Eq.(\ref{H1ch}). Since the Hamiltonian is quadratic we can easily evaluate all the equilibrium correlation functions at finite temperature, paying attention to the different time orderings that appear as a result of the two branches of the Keldysh contour. The results for finite temperature is:
\begin{widetext}
\begin{align} \label{Ctt}
&\overline{C^\theta(x,t)}= \frac{g}{4}\left[ \log \left(\frac{\beta v_L}{\pi \delta} \sinh \left(\frac{\pi(x+v_Lt -i\delta)}{\beta v_L}\right)\right) +  \log \left(\frac{\beta v_R}{\pi \delta} \sinh \left(\frac{\pi(x-v_Rt + i\delta)}{\beta v_R}\right)\right) + (x\rightarrow -x, t\rightarrow -t)\right];\\
&R^{\theta}(x,t)=-\frac{\pi}{2}g\left[ \Theta(x)\Theta(t-\frac{x}{v_R})+\Theta(-x)\Theta(t-\frac{|x|}{v_L})\right]
\end{align}
\end{widetext}

where $\delta$ is a short distance cutoff, $v_{R/L}=v_F/g \pm u $, and $\Theta(x)$ is the step function. As mentioned in Ref. \onlinecite{RecherEtAl}, it is important to remember that the step functions are not infinitely sharp, and have a transition width of order $a$, the cutoff. The functions $\overline{C^\phi(x,t)}$ and $R^\phi(x,t)$ are obtained from $\overline{C^\theta(x,t)}$ and $R^\theta(x,t)$ by replacing the prefactor $g$ with $\frac{1}{g}$. The function $C^{\theta\phi}$ is given in Eq. (\ref{Ctp}), and:

\begin{align}
&R^{\theta\phi}(x,t)=  \\
&-\frac{\pi}{2} \Theta(t) \left[ \Theta(x)\Theta(x-v_Rt)-\Theta(-x)\Theta(|x|-v_Lt) \right]. \nn
\end{align}
The currents are expressed in integrals over complicated combinations of such correlation functions. For example, the coherent part of the current, given by Eq. (\ref{Ico}), involves the following combinations:

\begin{align} \label{Cco}
&C_{co}(L,t)=2C^{\theta\phi}(L,t)-2C^{\theta\phi}(-L,t) \\
&+2\overline{C^\theta(0,t)}-2\overline{C^\theta(L,0)}+\overline{C^\theta(L,t)}+\overline{C^\theta(-L,t)}\nn \\
&+2\overline{C^\phi(L,0)}-2\overline{C^\phi(0,t)}+\overline{C^\phi(L,t)}+\overline{C^\phi(-L,t)} \nn
\end{align}
and
\begin{align} \label{Rco}
&R_{co}(L,t)= R^{\theta\phi}(L,t)-R^{\theta\phi}(-L,t) \\
&+R^{\theta}(0,t)+\frac{1}{2}R^{\theta}(L,t)+\frac{1}{2}R^{\theta}(-L,t)  \nn \\
&-R^{\phi}(0,t)+\frac{1}{2}R^{\phi}(L,t)+\frac{1}{2}R^{\phi}(-L,t) \nn
\end{align}

The corresponding functions for the incoherent current are :
\begin{align} \label{Cinco}
&C_{inco}^{\pm}(L,t)=\pm \left( 2C^{\theta\phi}(L,t)-2C^{\theta\phi}(-L,t)\right) \\
&+2\overline{C^\theta(0,t)}-2\overline{C^\theta(L,0)}+\overline{C^\theta(L,t)}+\overline{C^\theta(-L,t)}\nn \\
&+2\overline{C^\phi(L,0)}+2\overline{C^\phi(0,t)}-\overline{C^\phi(L,t)}-\overline{C^\phi(-L,t)} \nn
\end{align}
and
\begin{align} \label{Rinco}
&R_{inco}(L,t)= R^{\theta}(0,t)+\frac{1}{2}R^{\theta}(L,t)+\frac{1}{2}R^{\theta}(-L,t)  \nn \\
&+R^{\phi}(0,t)-\frac{1}{2}R^{\phi}(L,t)-\frac{1}{2}R^{\phi}(-L,t). \nn
\end{align}

In a Carbon nanotube there are four channels, rather than a single one. In the non-interacting case, $g=1$, all these channels are independent and we would recover the results of the single channel. Equations (\ref{Cco}) and (\ref{Rco}) still apply for this case. When $g\neq 1$, the different channels are coupled through the interaction, and we must find the correct combinations of the fields $\theta_{i\sigma}$ and $\phi_{i\sigma}$ which decouple and therefore diagonalize the Hamiltonian. These combinations are discussed in Appendix \ref{Diagonalization}. This change of basis is in general a function of $u/v_F$ and $g$, and it mixes the $\theta$ and $\phi$ fields, which in turn complicates the functions $C_{co}$ and $R_{co}$ further. Luckily, the interactions in Carbon nanotubes are strong, $g\approx 0.3$, and in that range, the change of basis is very close to the usual spin/node symmetric/antisymmetric change of basis. If we approximate the diagonalizing fields by these symmetric/antisymmetric combinations, then equations (\ref{Cco}) and (\ref{Rco}) would apply provided we make the following substitutions:
\be \label{subs}
\overline{C^{\theta}(x,t)}\rightarrow \frac{1}{4}\sum_{j=1..4} \overline{C^{\theta_j}(x,t)}
\ee
Where each $\theta_j$ has a different set of values for $v_R$, $v_L$ and $g$ to be used in Eq.(\ref{Ctt}). The fields $\theta_1$ and $\theta_2$ correspond to the spin asymmetric combinations, which decouple from the interaction, and hence have $g=1$, and velocities $v_R=v_F\pm u$ and $v_L=v_F \mp u$. The fields $\theta_3$ and $\theta_4$ both have the same left and right mover velocities, $v_3$ and $v_4$ respectively, given by Eq. (\ref{velocities}), and interaction parameters $1$ and $g$, respectively.

\bibliography{Sagnac}

\end{document}